%% file: dernier.tex
\documentstyle[epsf]{europhys}
\input euromacr

\input epsf
\begin{document}

\euro{}{}{1-$\infty$}{1999}
\Date{July 7, 1999}
\shorttitle{L. F. CUGLIANDOLO et al. FDT violations in spin-glasses}

\title{A Search for Fluctuation-Dissipation Theorem \\ Violations in
 Spin Glasses from Susceptibility Data}
\author{L. F. Cugliandolo\inst{1}, D. R. Grempel\inst{2}
	\footnote{On
       leave from Service de Physique Statistique, Magn\'{e}tisme et
       Supraconductivit\'{e}, DRFMC, F-38054 Grenoble, France.},
       J. Kurchan\inst{3},
       \And E. Vincent\inst{2}
}
\institute{
     \inst{1} Laboratoire de Physique Th\'eorique 
              de l'\'Ecole Normale Sup\'erieure de Paris,
             24 rue Lhomond, F-75231 Paris  
              and Laboratoire de Physique Th\'eorique  et Hautes \'Energies, 
              Univ. de Paris VI and VII
              Tour 24, 5\`eme \'etage, 4 Place Jussieu, F-75252 Paris, France\\
     \inst{2} Service de Physique de l'\'Etat Condens\'e, CEA-Saclay, 
              F-91191 Gif-sur-Yvette, France\\
     \inst{3} Laboratoire de Physique de l'\'Ecole Normale 
              Sup\'{e}rieure de Lyon, 
              46 All\'ee d'Italie, F-69364 Lyon, France\\
}
\maketitle
\begin{abstract}
We propose an indirect way of studying  
the fluctuation-dissipation relation in spin-glasses
that only uses available susceptibility data. It is based 
on a dynamic extension of the Parisi-Toulouse 
approximation and a Curie-Weiss treatment of the average
magnetic couplings. We present the results of the analysis of 
several sets of experimental data obtained from various samples.
\end{abstract}

{\it Introduction.}---
The fluctuation-dissipation theorem (FDT) relates 
the response of a magnetic system
to the magnetization correlation function 
at equilibrium. In its
integrated form FDT states that
\begin{equation}
\chi(t,t_w) = \frac{1}{T} (q_d - C(t,t_w))
\; ,
\label{FDT}
\end{equation}
where the response to an applied  field $h$ held constant from a
 waiting-time
$t_w$ up to $t$  and the correlation are defined as
\begin{eqnarray*}
\chi(t,t_w)\equiv\delta \langle m(t)\rangle/\delta h|_{h=0}
\; , 
\;\;\;\;\;\;\;\;\;
C(t,t_w)\equiv\langle m(t) m(t_w)\rangle
\; ,
\end{eqnarray*}
and $q_d$ is the long-time limit of $C(t,t)$. 

Glassy systems are out of equilibrium and FDT 
does not apply,  as  shown in particular
in solvable models with infinite range interactions.
Otherwise stated, there are effective temperatures (different
from the bath-temperature) at play in aging systems~\cite{Cukupe}. 
A simple modification of Eq.~(\ref{FDT}) consists in 
proposing that, for large $t_w$,  the susceptibility still
 depends on $t$ and $t_w$  {\it  only} through $C$, 
{\it i.e.}
\begin{equation}
\chi(t,t_w) = \chi(C(t,t_w))
\end{equation}
where $\chi(C)$ is a system-dependent function \cite{Cuku1}.
The latter may be obtained from 
a plot of $\chi(t,t_w)$ against $C(t,t_w)$
using $t\geq t_w$ as a parameter.
This curve is known analytically for several 
mean-field  spin-glasses \cite{review}
such as the Sherrington-Kirkpatrick (SK) model. Models with finite 
range interactions have been studied numerically 
by several groups \cite{numerics2,Juan3DEA} who obtained  results 
that are qualitatively similar to the mean-field 
ones. These numerical results  must however be taken 
with caution since the times that can be reached
in simulations are relatively short.

In principle, $\chi$ and $C$ should be determined experimentally by
measuring independently noise correlations and susceptibilities. This
procedure has been recently used to investigate FDT violation in
structural glasses~\cite{Grigera}.  For spin-glasses, noise
measurements along the lines of the early work of Ref.~\cite{oldies}
are under way: their outcome will provide a most stringent test for
spin-glass theories.

In this paper, we shall assume that violations of FDT do occur below
$T_g$, and propose an {\it indirect} method for the determination of
the $\chi$ vs $C$ curve using {\it available} experimental data. This
construction gives a first glimpse of the form of this curve. It is
based on some assumptions, notably a dynamic extension of the
Parisi-Toulouse (PaT) approximation~\cite{PaT}, that hold for some
spin-glasses.

We start by noticing that, in all solvable spin-glass models, below
$T_g$, $\chi(C)$ is a piecewise function \cite{review}:

\begin{eqnarray}
\chi(C) 
&=&
\left \{
\begin{array}{rcl}
\frac{1}{T} \left( q_d-C \right) \;\;\;& \mbox{if} & q \leq C < q_d
\; \ \ \ {\rm (FDT \ regime)},
\\
\chi_{\sc ag}(C) + \frac{1}{T} \left( q_d-q \right) \;\;\;& \mbox{if} & q_0 < C < q
\; \ \ \ {\rm (aging \ regime)},
\end{array}
\right.
\label{brokencurve}
\end{eqnarray}
where the dynamical Edwards-Anderson parameter, $q$,
and the minimal correlation in an applied magnetic field $H$, $q_{0}$, 
are defined as 
\begin{equation}
q\equiv\lim_{t-t_w\to\infty} \lim_{t_w\to\infty} C(t,t_w)
\; ,
\;\;\;\;\;\;
q_0\equiv \lim_{t\to\infty} C(t,t_w)
\; ,
\end{equation}
the latter vanishing in zero applied field.
In the SK model, $\chi_{\sc ag}(C)$ is 
decreasing  and has a downwards curvature. Numerical results indicate that, 
at least within the  
simulation times, the shape of the curve for the 3D 
Edwards-Anderson (EA) model is similar.

The dynamical version of the PaT hypothesis \cite{Juan3DEA,PaT} (see also 
Ref.~\cite{Cuku3}) consists of the 
following two assertions: (i) $\chi(C)$ is independent of 
$T$ and  $H$ in the aging regime and (ii) $q$ and $q_0$ only depend 
on $T$ and $H$, respectively. We shall moreover  assume 
that  this approximation is good even at finite times (see Discussion). 

The near temperature-independence of $\chi(C)$ in the aging regime 
has been checked numerically for the 4D EA model in Ref.~\cite{Juan3DEA}.
No checks are available for the 3D case. The validity of
this approximation for the experimental systems will be discussed below.
It will be seen that the PaT approximation allows us to estimate
 the $C$-dependence of the susceptibility  
using {\it exclusively} response results, thus circumventing
the difficulties inherent to noise measurements.

\begin{figure}[htbp]
\vskip 2.0cm
\begin{center}
   \leavevmode 
   \epsfysize=4.3cm\epsfbox{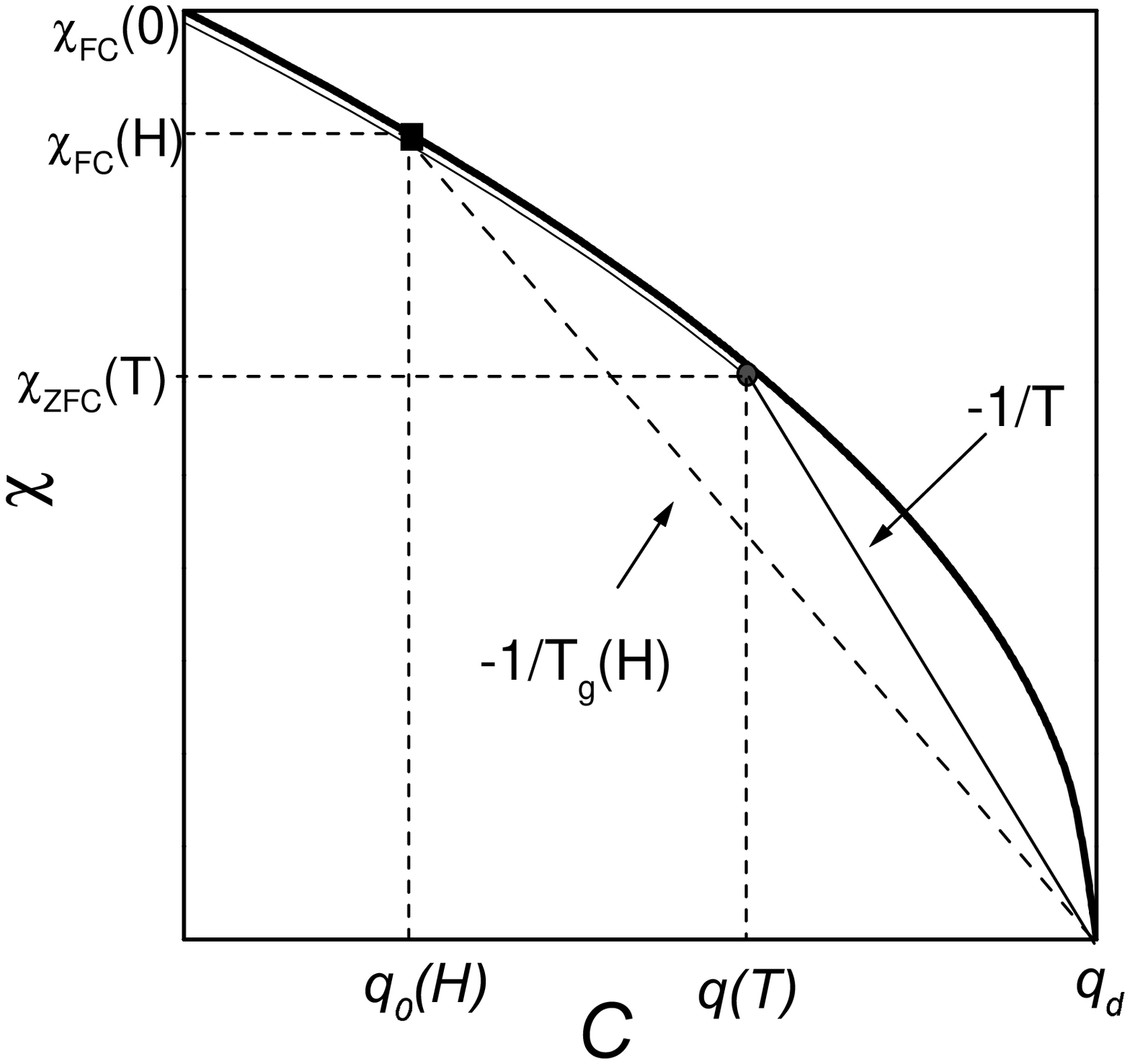}
    \hspace*{4.7mm}
    \epsfysize=4.3cm\epsfbox{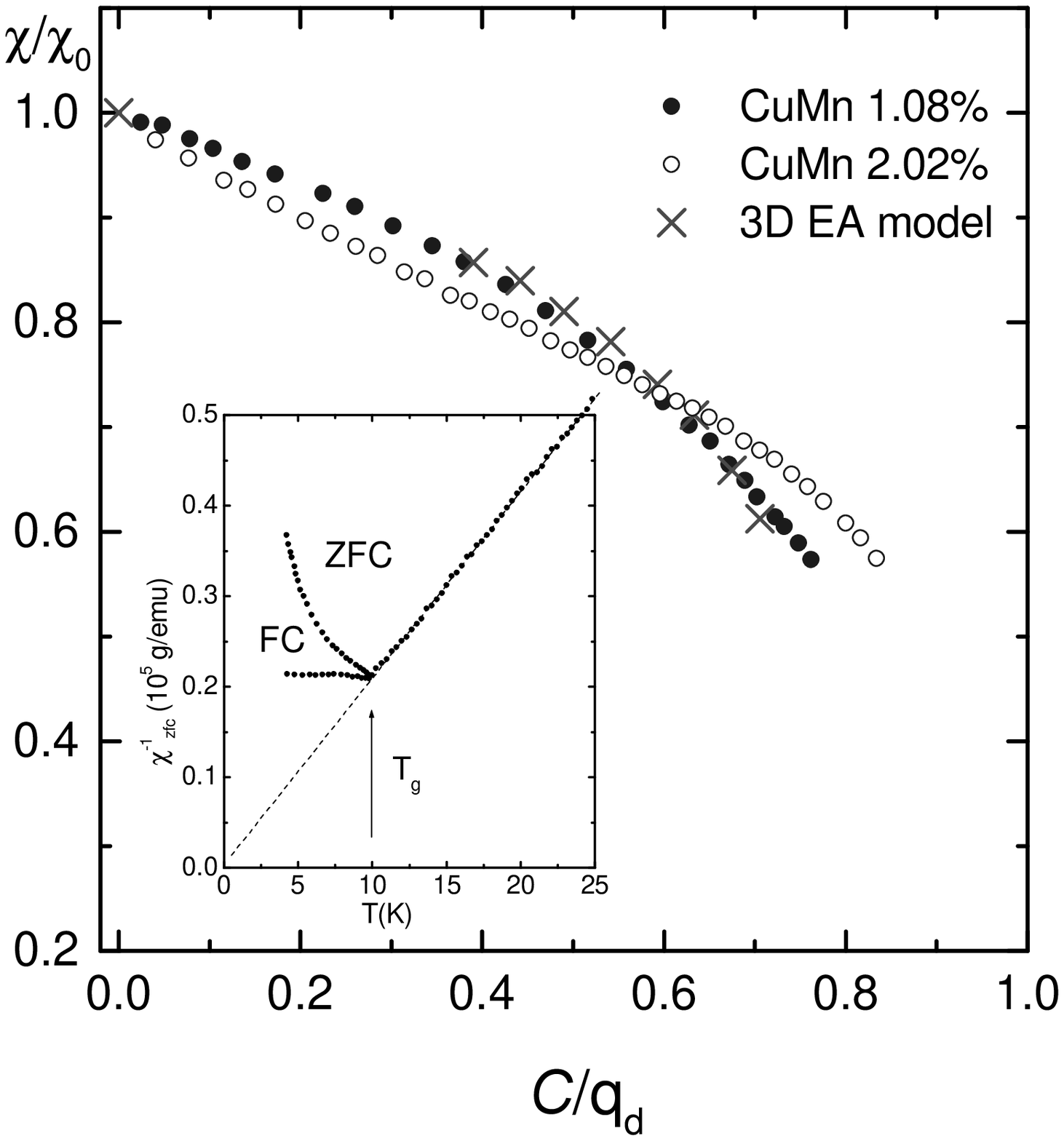}
  \end{center}
\vspace{.2cm}
  \hbox to \textwidth{\vspace*{5mm}\hfil Fig.\ \ref{sketch}\hfil\hfil 
  Fig.\ \ref{nagata}\hfil}
  \caption{\label{sketch}
     Sketch of the $\chi$ vs $C$ plot. The thick curve represents the
 master curve $\tilde{\chi}(C)$ that, within the PaT approximation, is
 temperature and field independent.  The thin straight line has slope
 $-1/T$ ($T<T_g$) and represents the first line in
 Eq.~(\ref{brokencurve}). The dashed straight line has a slope
 $-1/T_g(H)$ and joins $(q_d,0)$ to $(\chi_{\sc fc}(H),q_0(H))$.
       } 
  \caption{\label{nagata} 
    $\tilde{\chi}(C)$ plot for CuMn at 1\% and 2\%. Data are taken
  from reference \cite{Nagata}. The vertical axis is normalized by the
  susceptibility at the critical temperature in zero field
  ($\chi_0$). The horizontal axis is normalized by $q_d={\cal C}$.
  The crosses are numerical results for the 3D EA model
  \cite{Juan3DEA}. The inset shows the inverse ${\sc fc}$ and ${\sc
  zfc}$ susceptibilities as functions of temperature.
  }
\end{figure}

Our strategy is to use data taken under $T$ and $H$ conditions 
 such that 
 the system is at the limit of validity of FDT, {\it i.e.}
 $C(t,t_w)=q$. The point $\{ q,\chi(q)\}$ is the intersection between
 the straight part (FDT regime)
 and the curved part  (aging regime) of
 $\chi(C)$ (cf. Eq.\ \ref{brokencurve}). The
 locus of the points obtained varying $T$ and $H$ spans a {\it master
 curve} $\tilde{\chi}(C)$ which, by the PaT hypothesis, is field and
 temperature independent. The method of construction is explained
 below and 
illustrated 
in Fig~\ref{sketch}.

The susceptibility at the limit of the FDT regime corresponds to:
\begin{eqnarray}
%\chi^{\sc th}_{\sc zfc}(T,H) 
%&\equiv& 
\chi(q) = \lim_{t-t_w\to\infty} \lim_{t_w\to\infty}
\chi(t,t_w) = \frac{1}{T} (q_d - q)
\label{chizfc}
\;.
\end{eqnarray}
We have approximated this limit by using
 susceptibility data of three 
different types taken from the literature.

\underline{Frequency dependent measurements}.

In {\it ac}-susceptibility measurements, a small {\it ac}-field of fixed 
frequency $\omega$ is applied. 
The in-phase susceptibility $\chi'(\omega,t_w)$ 
is recorded as a function of temperature. 
For frequencies in the range $\omega\stackrel{>}{\sim}$ 1Hz,  the long
waiting-time limit $\omega t_w \gg 1$ 
 is approached within
the measurement time.
Then we can estimate the limit of zero frequency by extrapolation 

\begin{equation}
\chi'(0,\infty)
\equiv
\lim_{\omega\to 0} \lim_{t_w\to\infty} 
\chi'(\omega,t_w) = 
\frac{1}{T}(q_d-q(T))
\; .
\end{equation}

The master curve $\tilde{\chi}(C)$ is obtained by joining the points
$\left\{C=q_d-T \chi'(0,\infty) \, ; \; \tilde\chi=\chi'(0,\infty)
\right\}$ using $T$ as parameter.

\underline{Field cooled measurements}.  $M_{\sc fc}$ is measured by
cooling the sample in a constant magnetic field.  Below $T_g$,
$\chi_{\sc fc}=dM_{\sc fc}/dH $ rapidly reaches an asymptotic value
(see however \cite{Nordblad-new}).

In some spin-glasses like CuMn \cite{Nagata} $\chi_{\sc fc}$ is nearly
temperature-independent below $T_g$, $\chi_{\sc fc}(T,H) \sim
\chi_{\sc fc}(H)$ as required by PaT.  However, this does not hold in
most systems for small fields and near $T_g$ where a cusp in
$\chi_{\sc fc}$ appears.  The $\chi_{\sc fc}$ data may be used where
both FDT and PaT hold: on the critical line (assuming there is at
least a transient one, an issue discussed below).  The parameter is
now $H$ and the master curve is spanned by the points
$\left\{C=q_d-T_g(H) \chi_{\sc fc}(H) \, ; \; \tilde\chi=\chi_{\sc
fc}(H) \right\}$.

\underline{Zero-field cooled measurements}. 

In the absence of a sufficient amount of published {\it ac} or ${\sc fc}$ data, we
have resorted to include in the analysis data obtained in a zero-field
cooled procedure. 
The sample is 
quenched in the absence of a field from above $T_g$ down to
some low temperature. After a time $t_w$ (necessary for stabilization
of the temperature), 
a weak magnetic field $H$ is applied and $M_{\sc zfc}$ is immediately measured. Under
typical experimental conditions, the measurement time $t$ is
significantly shorter than $t_w$. We shall thus consider that
$\chi_{\sc zfc}\equiv M_{\sc zfc}/H$ is a good approximation 
to Eq.~(\ref{chizfc}).

Ideally, the same procedure should be repeated for each measurement
temperature. In practice, however, the magnetization $M_{\sc zfc}$ is
measured by increasing the temperature in steps from its initial
value. Although the two methods are not strictly equivalent, we
believe that the possible differences are of little consequence for
our conclusions. Therefore we shall consider that the whole
experimental $M_{\sc zfc}(T)$ curve yields an acceptable approximation
to Eq.~(\ref{chizfc}). Support for this point of view is given by a
comparison of both {\it ac} and ${\sc zfc}$ procedures on one sample
(see below).

The field-independence  
of $\chi_{\sc zfc}$ implied by PaT is in general well verified  
experimentally except for the  largest fields (see, for 
example, Fig.~1 of the first of Refs.~\cite{Ito}). The master curve is
determined as in the {\it ac}case.

Before turning to the analysis of the data,  
we notice that Eq.~(\ref{brokencurve}) cannot be 
applied to experimental data as it stands. 
Indeed, this equation is only valid for 
systems in which the exchange coupling 
averages to zero.
In real spin-glasses, however, this average is, in general, finite. 
This is reflected in the behavior of the high temperature susceptibility
that often obeys a Curie-Weiss law. The Curie-Weiss temperature 
$\theta$ may be a sizeable fraction of $T_g$. 
Within a mean-field approximation of the average coupling, however, 
Eq.~(\ref{brokencurve}) still holds for the response to the {\it total} 
field (applied plus internal). This amounts to the replacement
$
\chi \to \chi_{\sc meas}/(1 +\theta/{\cal C }\; \chi_{\sc meas})
$,
with $\chi_{\sc meas}$ the measured susceptibility and ${\cal C}$ 
the Curie constant. In some metallic spin-glasses, like CuMn 
at low concentration of the magnetic impurity \cite{Nagata}, 
as well as AuFe \cite{Canella} and AgMn \cite{orbach},
the Curie-Weiss law holds down to the transition temperature.
This is not the case in other spin-glasses where, due to progressive 
clustering, 
the paramagnetic behaviour 
deviates from a simple Curie-Weiss law close to the 
transition. 
When the Curie-Weiss law holds, 
a plot of the inverse susceptibility as function of 
temperature yields the values of ${\cal C}$ and $\theta$ and it is easy to show
that ${\cal C}=q_d$. 

\vspace{.2cm} 
{\it The analysis. ---} We have analysed data obtained
with the methods described above for three metallic systems: CuMn
\cite{Nagata} and AuFe \cite{Canella} for several concentrations, AgMn
at 2.6\% \cite{orbach}, and two insulating samples
CdCr$_{1.7}$In$_{0.3}$S$_4$ \cite{Lefloch,Vincent} and Fe$_{0.5}$
Mn$_{0.5}$TiO$_3$ \cite{Ito}.

For CdCr$_{1.7}$In$_{0.3}$S$_4$  
we used the $\chi'(\omega,t_w)$ data \cite{Vincent}, and extrapolated
to zero frequency assuming a
 power-law decay  $\chi'(\omega,\infty) \sim 
\chi'(0,\infty) + c_1 \omega^a$. In the cases of CdCr$_{1.7}$In$_{0.3}$S$_4$ 
and Fe$_{0.5}$ Mn$_{0.5}$TiO$_3$, we also used the ${\sc fc}$
data to check the consistency of our determination. 
The susceptibilities $\chi_{\sc zfc}$  were obtained 
for CuMn, AgMn, AuFe, CdCr$_{1.7}$In$_{0.3}$S$_4$
and Fe$_{0.5}$ Mn$_{0.5}$TiO$_3$ 
using $\chi_{\sc zfc}\approx M_{\sc zfc}/H$. 

In order to analyse the field-cooled data, it is necessary
to identify the transition temperature in the presence of an
applied field. This is done
by studying the onset of irreversibilities in the 
magnetization curves: 
we defined $T_g$ as the temperature where 
$M_{\sc irr} \equiv M_{\sc fc}-M_{\sc zfc} = 0$.
At high fields, when the relationship between $\chi_{\sc fc}$ and 
$M_{\sc fc}$ is non-linear, we have made polynomial fits 
of $M_{\sc fc}(H)$ to compute $\chi_{\sc fc}(H)=dM_{\sc fc}/dH$. 
For high enough fields, $M_{\sc fc}$ 
is independent of temperature supporting 
the PaT approximation. However, for some samples, 
a cusp in the low-field susceptibility may appear 
near the transition temperature     
and the determination of 
a $T$-independent $\chi_{\sc fc}$ becomes ambiguous. 
In these cases we have used the value 
$\chi_{\sc fc}$ at the critical temperature in order to ensure 
consistency with 
the alternative determination of $\tilde{\chi}(C)$ 
based on $M_{\sc zfc}$ measurements. 
On the whole, the construction using ${\sc fc}$ data
is more ambiguous than that using ${\sc zfc}$ or {\it ac} data.

Our analysis is most reliable for CuMn, 
a system in which the Curie-Weiss law 
 as well as the PaT approximation 
are very well verified.  Figure~\ref{nagata} shows the $\tilde{\chi}(C)$  
curve determined using the ${\sc zfc}$ data of Ref.~\cite{Nagata} for 
two concentrations, 1.08 \% and 2.02\%. There are no experimental 
points for $C/q_d>0.8$ that correspond to rather low temperatures.   
We know however that $\tilde{\chi}(C)$ tends to zero as 
$C \rightarrow q_d$ since $\chi_{\sc zfc}(T=0)=0$.
In addition, the slope $d\tilde{\chi}/dC$ should be infinite at $C=q_d$
so that $q=q_d$ only at $T=0$. 
The validity of the hypotheses can be judged by the 
inset of Fig.~\ref{nagata} where we show the temperature 
dependence of the inverse susceptibiliy for the 1.08\% compound. 
A  Curie-Weiss law with $\theta \approx 0$ holds accurately for all 
$T \ge T_g$. The $T$-independence of 
$\chi_{\sc fc}$ required by the PaT approximation is also 
well verified below the transition. The same is true for 
the 2.02\% sample.

\begin{figure}[htbp]
\vskip 1.5cm
\begin{center}
   \leavevmode 
   \epsfysize=4.4cm\epsfbox{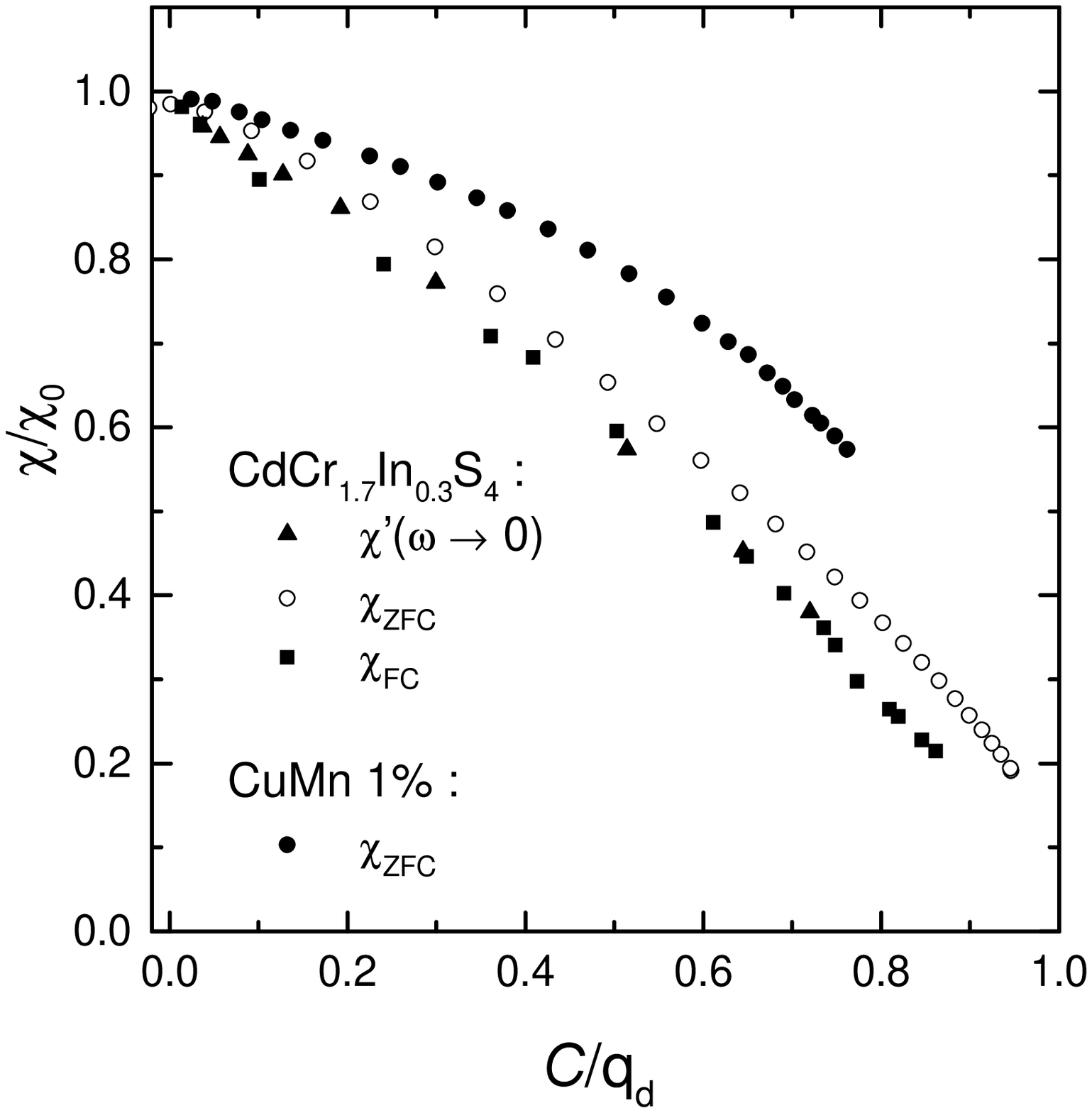}
    \hspace*{4.7mm}
    \epsfysize=4.4cm\epsfbox{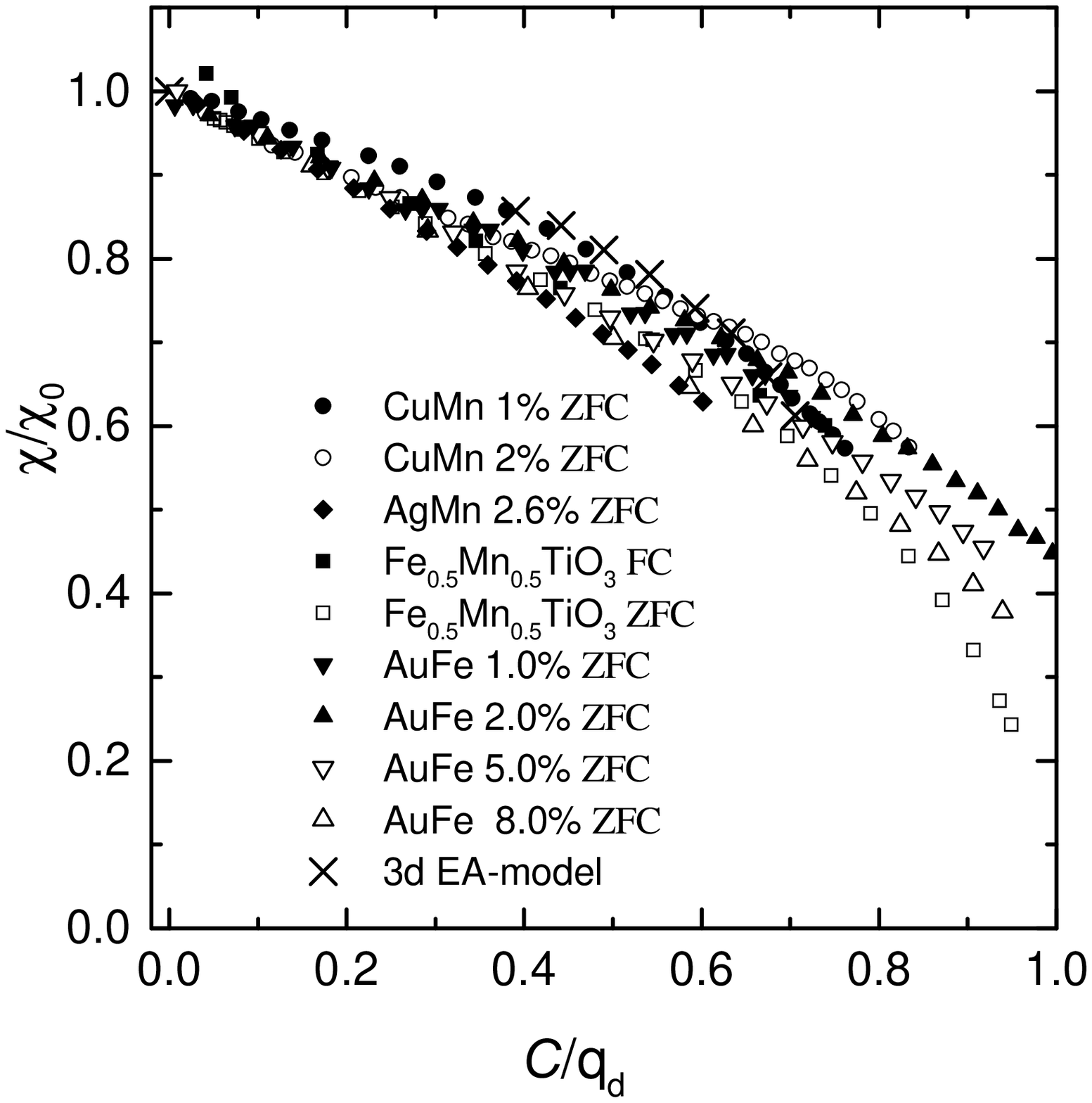}
  \end{center}
%\vspace{.4cm}
  \hbox to \textwidth{\vspace*{5mm}\hfil Fig.\ \ref{eric}\hfil\hfil 
  Fig.\ \ref{todas}\hfil}
  \caption{\label{eric}
     $\tilde{\chi}(C)$ curves for CdCr$_{1.7}$In$_{0.3}$S$_4$ determined from 
    ${\sc fc}$, ${\sc zfc}$ and {\it ac} susceptibilities. We show for
  comparison the ${\sc zfc}$ data for CuMn 1.08\%.   
  }
  \caption{\label{todas} 
      $\tilde{\chi}(C)$ curves for some of the samples studied. ${\sc
    zfc}$ data for CuMn at 1.08\% and 2.02 \%; AuFe at 1.0\%, 2.0\%,
    4\% and 8.0\%; AgMn at 2.6\% and Fe$_{0.5}$ Mn$_{0.5}$TiO$_3$;
    ${\sc fc}$ data for Fe$_{0.5}$ Mn$_{0.5}$TiO$_3$.
  }

\end{figure}

For comparison, we also show in Fig.\ \ref{nagata} the curve
$\tilde{\chi}(C)$ for the 3D EA model, at $T=0.7 (<T_g)$ and $H=0$,
obtained numerically in Ref.~\cite{Juan3DEA}.  The agreement between
the numerical results and the experimental data for the 1.08\% sample
is remarkable. It may be fortuitious, however, since the results for
the 2.02\% sample deviate from it. In fact, one must note that
$\tilde{\chi}(C)$ is not a universal function. For example, it depends
on the details of the Hamiltonian (Heisenberg, Ising and, in general,
the level of anisotropy) even at the mean-field level.  Thus, there is
no reason to expect universality in real systems.

The data for CdCr$_{1.7}$In$_{0.3}$S$_4$ \cite{Lefloch,Vincent}
obtained using the three different techniques described above are
shown in Fig.~\ref{eric}.  It can be noticed that the ${\sc fc}$ and
{\it ac} results are very similar. The ${\sc zfc}$ data are somewhat
higher (probably due to the fact that the large $t_w$ condition is not
as well full filled as in the other cases), but the agreement with the
other determinations remains acceptable.  Notice that the
$\tilde{\chi}(C)$ curve obtained for this compound is quite different
from that corresponding to CuMn.

In Fig.~\ref{todas} we collect results for all the other samples. 
As expected, the curves do not 
fall on a universal curve, but their shape is  similar. 

\vspace{.2cm}

{\it Discussion.}---
Finally,  let us clarify an important point. 
In several situations, such as 
2D Ising and 3D Heisenberg \cite{Kawamura} and, perhaps,
3D Ising spin glasses under a magnetic field, 
no true spin-glass phase is expected. However, 
for still relatively long times the system 
remains below a slowly time-dependent {\it pseudo} 
de Almeida-Thouless (AT) line~\cite{deAT}: it 
ages and behaves as a true (out of equilibrium) glass
with a non-trivial $\chi(C)$ that would  
eventually become a straight line with slope $-1/T$.
In this paper we explored the consequences of the 
stronger assumption that PaT is a  
good approximation below the pseudo AT-line if all quantities
involved belong to the same epochs. 

Another important issue is the asymptotic ($t_w\to\infty$) 
form of the $\chi_{\sc ag}(C)$ curve. 
Even if the system never equilibrates,
the $\chi_{\sc ag}(C)$ curve may still be a very slowly
varying function of $t_w$, eventually
reaching a form different from that  observed 
experimentally. We are not in a position to discard this possibility.

It has been recently shown that, under certain hypotheses,
the slope of the dynamic $\chi(C)$, for an 
infinite system in the large-$t_w$ limit coincides
with the static $x(q)$ as defined by the probability of 
overlaps of configurations
taken with the Gibbs measure \cite{Franz}.  
 Since we do not address here the issue as to whether 
the AT-line and the PaT approximation survive
beyond experimental times
 we cannot make any statements concerning
the relation of this dynamical function to the corresponding 
equilibrium Parisi function.

In conclusion we have presented an approximate determination of
the $\tilde{\chi}(C)$ curve characterising the deviations from equilibrium 
of spin-glasses through the violations of FDT, assuming they exist. 
This construction does not replace a true determination
via simultaneous measurements of susceptibility and noise correlation,  
as the ones in Ref.~\cite{Grigera,oldies}, 
but it yields some insight into how this curve might look in reality.  

\vspace{.2cm}
\centerline{***}
\vspace{.2cm}

We wish to thank D. Sherrington for early discussions on this subject as
well as H. Aruga-Katori, N. Bontemps, A. Ito, H. Takayama, A. Tobo 
and H. Yoshino. We also thank J. Hammann and M. Ocio for a critical
reading of the manuscript. 
LFC, DRG and EV thank the Monsbusho contract for partial
financial support. JK  was partially supported by the 
`Programme Th\'ematique Mat\'eriaux', R\'egion Rh\^one-Alpes.

\end{document}

%% file: euromacr.tex
\def\And{{\rm and\ }}

\newif\ifboo \boofalse